\documentclass[ floatfix,secnumarabic, amssymb, nobibnotes, fleqn,10pt]{wlscirep}

\usepackage{mathrsfs}

\usepackage{amssymb, mathtools, amsmath}

\usepackage{bm}

\usepackage{graphicx}
\usepackage{tikz}
\usetikzlibrary{arrows, shapes, positioning, calc}
\usepackage{quantikz}

\usepackage{quantikz}

\usepackage[utf8]{inputenc}
\usepackage[T1]{fontenc}

\title{Quantum Reservoir GAN}

\author[1,*]{ Hikaru Wakaura }

\affil[1]{ QuantScape Inc. QuantScape Inc., 4-11-18, Manshon-Shimizudai, Meguro, Tokyo, 153-0064, Japan }

\affil[*]{ hikaruwakaura@gmail.com }


\begin{abstract}

Quantum machine learning is known as one of the promising applications of quantum computers.
Many types of quantum machine learning methods have been released, such as Quantum Annealer, Quantum Neural Network, Variational Quantum Algorithms, and Quantum Reservoir Computers.
They can work, consuming far less energy for networks of equivalent size.
Quantum Reservoir Computers, in particular, have no limit on the size of input data.
However, their accuracy is not enough for practical use, and the effort to improve accuracy is mainly focused on hardware improvements.
Therefore, we propose the approach from software called Quantum Reservoir Generative Adversarial Network (GAN), which uses Quantum Reservoir Computers as a generator of GAN.
We performed the generation of handwritten single digits and monochrome pictures on the CIFAR-10 and Fashion-MNIST datasets.
As a result, Quantum Reservoir GAN is confirmed to be more accurate than Quantum GAN, Classical Neural Network, and ordinary Quantum Reservoir Computers.

\end{abstract}

\begin{document}
 
\flushbottom
\maketitle
%
%
\thispagestyle{empty}

\section{Introduction}\label{1}

Quantum computers are promised to solve some problems that classical computers are not able to solve since Lichard P Feynman proposed in 1982 \cite{feynman_simulating_1982}.
Some algorithms that can solve problems much faster than classical computers are reported by some groups, such as Grover algorithm \cite{2003quant.ph..1079L}, Quantum Fourier transform \cite{PRXQuantum.4.040318} and its applications \cite{2024arXiv241004435I,2023arXiv230504908M}, and Shor algorithms \cite{2023arXiv230609122S}.
However, quantum computers must avoid any types of noise to keep the fidelity of calculations rather than implementation \cite{2024npjQI..10...18D} of required gates and circumvent noise is hard due to the lack of fault-tolerant architecture; hence, today's openly available quantum computers are mainly Noisy Intermediate-Scale Quantum(NISQ) devices.

Meanwhile, many methods of quantum calculation that work on  NISQ devices, for example Variational Quantum Algorithms  \cite{Kassal2011,McClean_2016,Grimsley2019,2019arXiv190608728P,2021arXiv210501141W,2021arXiv210902009W}, Quantum Annealer \cite{d-wave2011}, Measurement Based quantum computer \cite{2009NatPh...5...19B}, Continuous Photon Quantum Computer  \cite{2024PhRvA.110a2607A}, and Quantum Reservoir Computers \cite{Fujii2020}.
Quantum Reservoir Computers (QRC) are a quantum version of Reservoir Computers.
Reservoir Computers take advantage of the hysteresis effects of materials.
They can memorize the input and reflect it on the output; hence, they can take the place of Neural Network algorithms.
QRC uses quantum phenomena as a nonlinear phenomenon in reservoirs.
They are implemented on Neutron Magnetic Resonance (NMR) \cite{Fujii2020}, photon qubits \cite{Nerenberg:25}, quantum dots \cite{Tate:24}, transmon \cite{dudas_quantum_2023} and so on \cite{2025CSF...19516289L}.
They are promised that because they can work with a small energy cost for an equivalent size of Neural Networks.
Hence, they have studied to apply for various calculations such as sound classification, prediction of waves, feedback control of robots, and solving partial differential equations \cite{yan_emerging_2024}.
Besides, industrial uses are surveyed by some groups \cite{2025arXiv250513287F}.

However, the accuracy of QRC is not high enough for practical use for small numbers of qubits.
Some groups try to improve the accuracy by combining it with Neural Networks \cite{2024arXiv240702553K}, modifying the algorithm itself \cite{PRXQuantum.5.040325}, and so on \cite{Nerenberg:25}.
QRC has the room to embed the parameters to improve the accuracy.
Generative Adversarial Network (GAN) \cite{2014arXiv1406.2661G} and families \cite{2017arXiv170400028G,2015arXiv151106434R,2016arXiv161104076M} have been used for generating large-scale complex data such as music and pictures.
The scheme of this is the duel between two networks in which one distinguishes the genuine data from generated fake data, and the other generates elaborate fake data to mimic the discriminator.
This scheme brought a significant improvement in accuracy and quality of data generated by artificial intelligence, and many improved versions and applications are reported \cite{2017arXiv170400028G,2015arXiv151106434R,2016arXiv161104076M}.
Their schemes are available for QRC to improve the accuracy and generate larger data.

Therefore, we propose a novel  QRC embedded in a GAN called Quantum Reservoir GAN (QRGAN) that uses QRC as a generator of GAN.
We demonstrated the training and generation of single handwritten digits on the Optdigit data set and pictures on the CIFAR10 dataset \cite{Krizhevsky2009LearningML}.
As a result, we confirmed that QRGAN is the most accurate compared to Quantum GAN and classical GAN for the datasets.

Section \ref{1} is the introduction, Section \ref{2} describes the method in detail of QRGAN, Section \ref{3} describes the result of training and generating the pictures, Section \ref{5} is the discussion of our results,  and Section  \ref{7} is the concluding remark.

\section{Method}\label{2}
   
In this section, we describe the details of QRGAN. 
QRGAN is QRGAN that uses QRC as a generator, to say frankly.
GAN is one of the generative models that combines two networks or models in machine learning.
The discriminator distinguishes genuine data from fake data generated by the generator network, and the generator network makes the fake data pass the discriminator's judgment as genuine data.  
Both are optimized to accomplish the assigned task, and they are expressed as an equation,

\begin{equation} 
\label{eq_gan} 
\min_G \max_D V_{\text{\tiny GAN}}(D, G) = \mathbb{E}_{\bm{x} \sim p_{\text{data}}(\bm{x})}[\log D(\bm{x})] + \mathbb{E}_{\bm{z} \sim p_{\bm{z}}(\bm{z})}[\log (1-  D(G(\bm{z})))] 
\end{equation}

Then, $ D(\bm{x}) $ is the probability that the data $ \bm{x} $ is distinguished correctly for the boolean label of the data $  p_{\text{data}}(\bm{x}) $ on discriminator $ D $, and $ G(\bm{z}) $ is the data made by generator $ G $ from latent vector $ \bm{z} $, respectively.
The boolean label of the data $ \sim p_{\text{data}}(\bm{x}) $ becomes 0 if the data is fake and 1 if genuine, and $  p_{\bm{z}}(\bm{z}) $ is 1 for all cases.
In simple form, it is expressed as,

\begin{eqnarray}
V_{\text{\tiny GAN}}(D, G) &=& L _D + L _G \\
L _D &=& p \log D(\bm{x}) + (1-  p) \log D(G(\bm{z}))    \\
L _G &=&-  p \log (1-  D(G(\bm{z})))   \\ \nonumber
\end{eqnarray}

which $ p $ is the boolean label.
We generate the $ G(\bm{z}) $ by the Quantum Reservoir Computer using the latent vector $ z$  as noise.
QRC is one of the QML methods that makes the filter matrix from input data and the encoded result of the time propagation of the quantum system.
This method can learn and predict from data without an optimization method.
Implementation ways of this are various, hence we use the example of NMR.
The system has 1 input qubit and $N_q$ output qubits.

 The system is initialized by the noise and time propagation-like ansatz $ exp (-  i \sum ( \theta_j + z_j)  Y_j + \sum (\theta_{ N_q + j } + z_{ N_q + j })  Y_j Y_k) $   which the number of parameters is $ 2 N_q + 3 N_q (N_q-  1) / 2 $.
One of the input data for j- th sample is input as the gate $ I-  x_{ j, t } Z_0 $ and the system is propagated for a $ w $ periods along the parametric Hamiltonian $ exp (-  i \pi w (\sum \theta_j ^h  Z_ j + \sum \theta_{ N_q + j } ^h Y_j Y_k + \sum \theta_{ N_q  + N_q (N_q-  1) / 2 + j } ^h X_ j X_k )) $. 
$ V_{ j t , l } = { \langle \Psi_j ( t \delta t) \mid Z_l \mid \Psi_j ( t \delta t) \rangle } $ is sampled at the end of propagation and reset the 0-  th qubit where V is $ N_{ sample } T \times N_q + 1 $ sized matrix that $ V_{ j t , N_q + 1 } = 1 $.     
Then, $ N_{ sample } $, $ T $, $ N_q $, $ \Psi_j ( t \delta t) $ and $ \delta t $ are the number of samples, number of time frames, number of qubits, state of quantum system at j-th sample at time $ t \delta t $, and time frame equivalent to $ \pi w $, respectively.
We repeat the above processes, inputting the following input data $ x_{ j, t + 1 } $ T times.
After making the matrix $ V $, we assume the filter matrix $ W $ as, 
\begin{equation}            
W = V ^{-1 } y,      
\end{equation}             
with fixed teacher data $ y $ equivalent to input vector $ x $.    
The predicted result is $ \tilde { y } = V W $.   
$ \tilde { y } $ is $ G(\bm{z}) $ on eq. \ref{eq_gan}. 
We use stochastic gradient descent as an optimizer on both the discriminator and the generator, and the learning rates are 0.1 and 0.001, respectively.    
We train on one image at once on an iteration, each of which is generated and separated into 4 patches.  
The teacher data is a connected vector of two copies of each batch of genuine data, and we use under half of the generated batch to combine into fake data.
The number of samples to process on the QRC of an iteration is fixed to 1. 
$ N_q $ is 4 for all calculations by QRGAN.     
All calculations are performed using PyTorch and Google Colab.

\section{Result of numerical simulations}\label{3}

In this section, we describe the results of generating the numbers on the Optdigit dataset and the pictures on the CIFAR10 dataset.
First, we compared the results on QRGAN, Quantum GAN (QGAN), and Classical Neural Network (CNN) using loss functions of cross-entropy and that of Least Squares on 8 $ \times $ 8-sized handwritten 0 for 375 samples, respectively.
QGAN \cite{2021PhRvP..16b4051H} is simulated by PennyLane.

\begin{figure}

\includegraphics[scale=0.3]{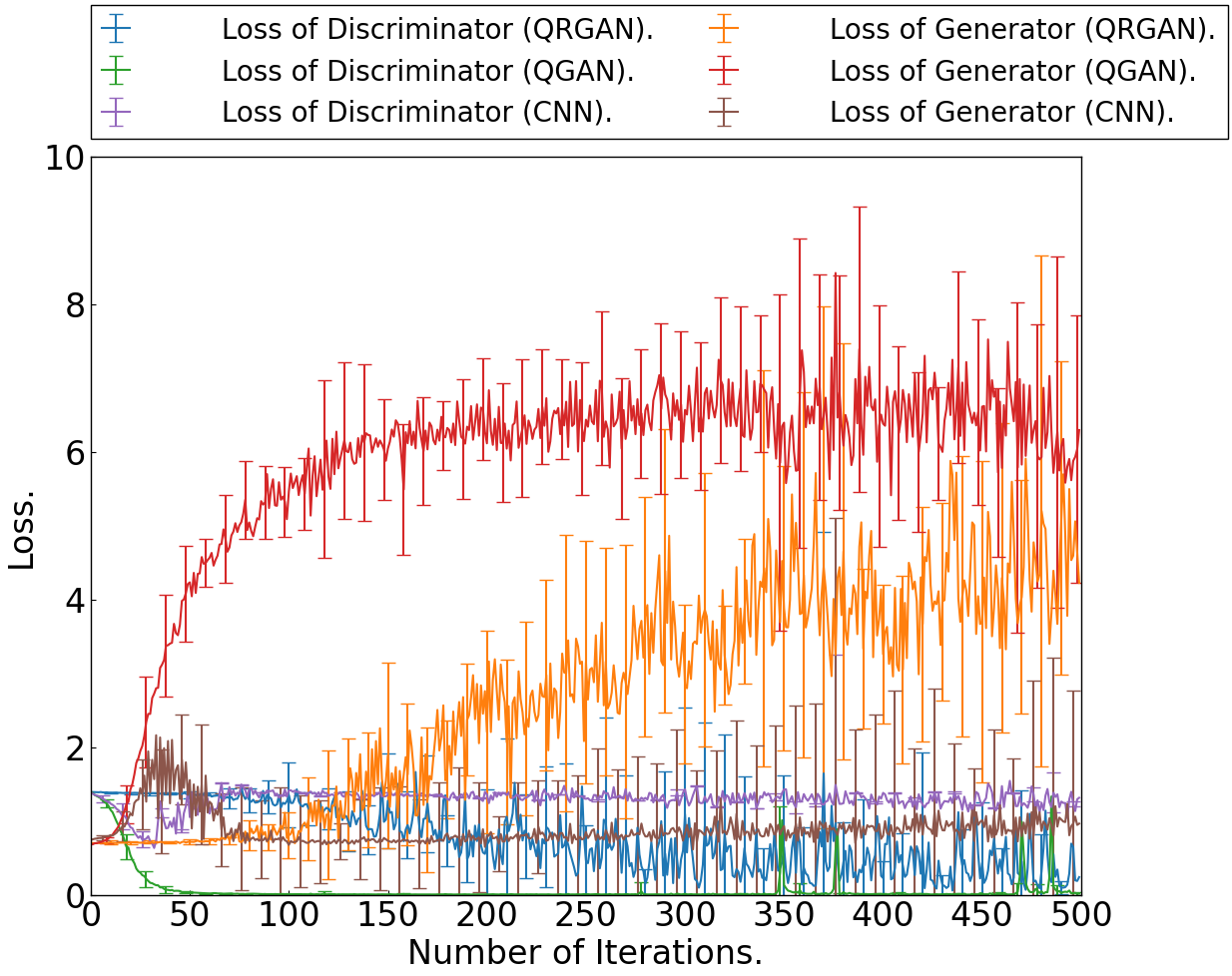}

\caption{ The number of iterations v.s. the average of the loss function for 10 attempts of the discriminator $ L_D$  and generator  $ L_G$   of QRGAN, QGAN, and CNN using cross-entropy as a loss function.
Error bars indicate the standard deviation from the average, and each bar is sampled every 10 points, moving 2 points from the previous data.
} \label{ comp 0 n }

\end{figure}

\begin{figure}

\includegraphics[scale=0.3]{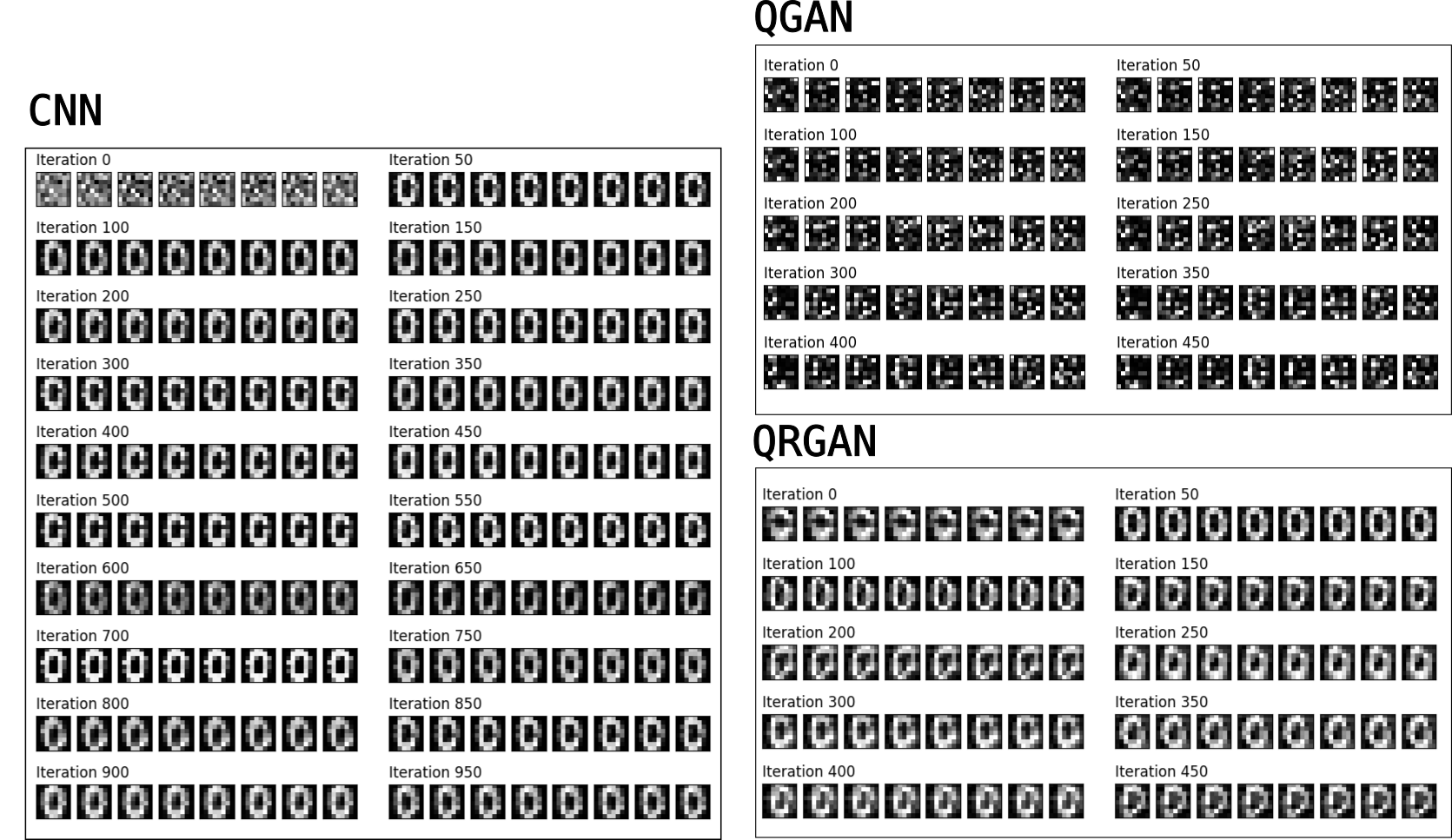}

\caption{ The generated picture of handwritten 0 by QRGAN, QGAN, and CNN using cross-entropy as a loss function for the number of iterations.
The picture by QRGAN is generated by fixed noise and input.  } \label{ gen 0 n }

\end{figure}

Number of qubit is 5 and the depth $ D $ is 6, and ansatz is $ \prod _{ j = 0 } ^{ N_q-  1 } R y (z_j) \prod _{ k   = 0 } ^{ D-  1 } \prod _{ j = 0 } ^{ N_q-  1 } R y (\theta_ { j, k }) \prod _{ j = 0 } ^{ N_q-  2 } C Z _{ j, j + 1 } $ for noise and parameters for each qubit.
We use under half of the distribution generated by the quantum process as the batch.
CNN is performed by a 3-layer network consisting of an input layer to expand 64 nodes to 128, a process layer to expand 128 nodes into 256, a process layer to expand 256 nodes into 786, and an output layer to compress 786 nodes to 64.
We show the average of the loss function for 10 attempts of the discriminator and generator of QRGAN, QGAN, and CNN for the number of epochs using cross-entropy in Fig.\ref{ comp 0 n }, and the generated picture of 0 in Fig.\ref{ gen 0 n }, respectively.
The loss function of the generator on QGAN is larger and rises faster than that of QRGAN on average, as shown in Fig.\ref{ comp 0 n }.
The loss function of the discriminator on QGAN depresses faster than that of QRGAN on average, too.
The whole process of both seems to have succeeded according to only loss functions, and QGAN is more accurate than QRGAN.

\begin{figure}

\includegraphics[scale=0.3]{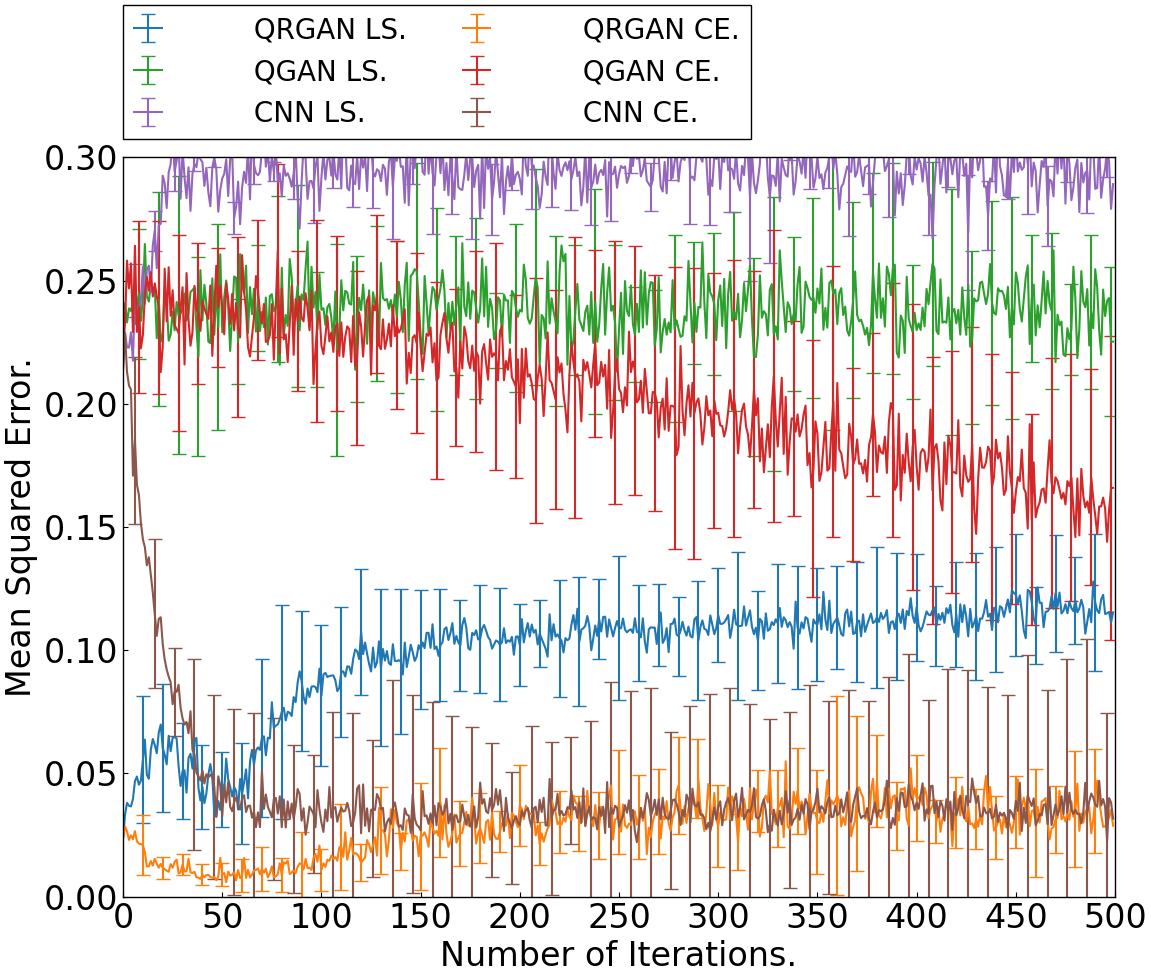}

\caption{
The number of iterations v.s. the average of the mean squared error (MSE) of QRGAN, QGAN, and CNN using the loss function of Least Squares GAN (LS) and cross-entropy (CE) as a loss function.}\label{ err }

\end{figure}

However, generated pictures of 0 of QRGAN are closer to the natural shape of 0 than those of QGAN, even for small numbers of epochs, according to Fig.\ref{ err }.
Besides, QRGAN generates the fakes closer to genuine data than CNN for about 100 iterations.
The accuracy of QRGAN surpasses the accuracy of QRC without optimization.

\begin{figure}

\includegraphics[scale=0.3]{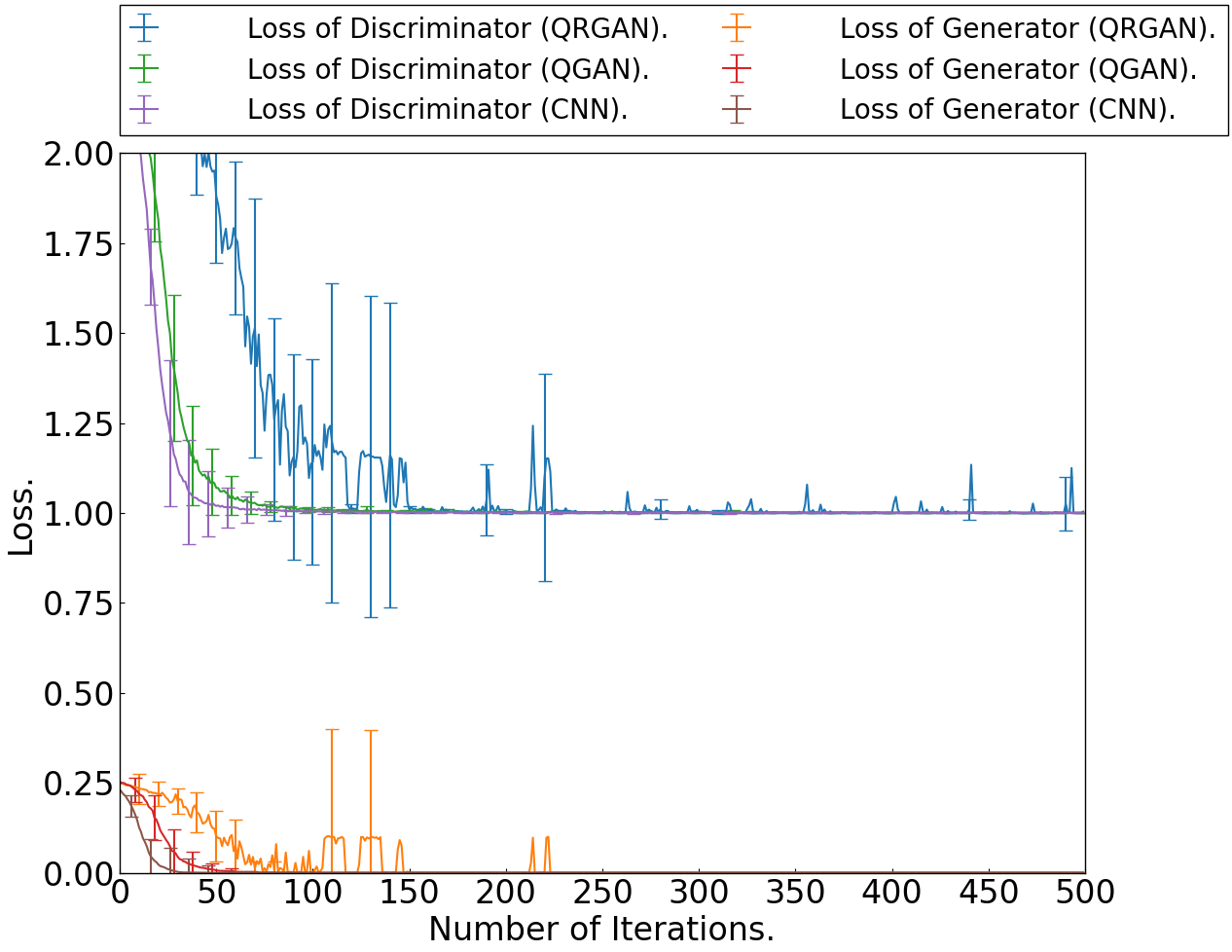}

\caption{The number of iterations v.s. the average of the loss function for 10 attempts of the discriminator $ L_D$  and generator  $ L_G$   of QRGAN, QGAN, and CNN using the loss function of Least Squares GAN as a loss function.  }\label{ comp 0 l }

\end{figure}

\begin{figure}

\includegraphics[scale=0.3]{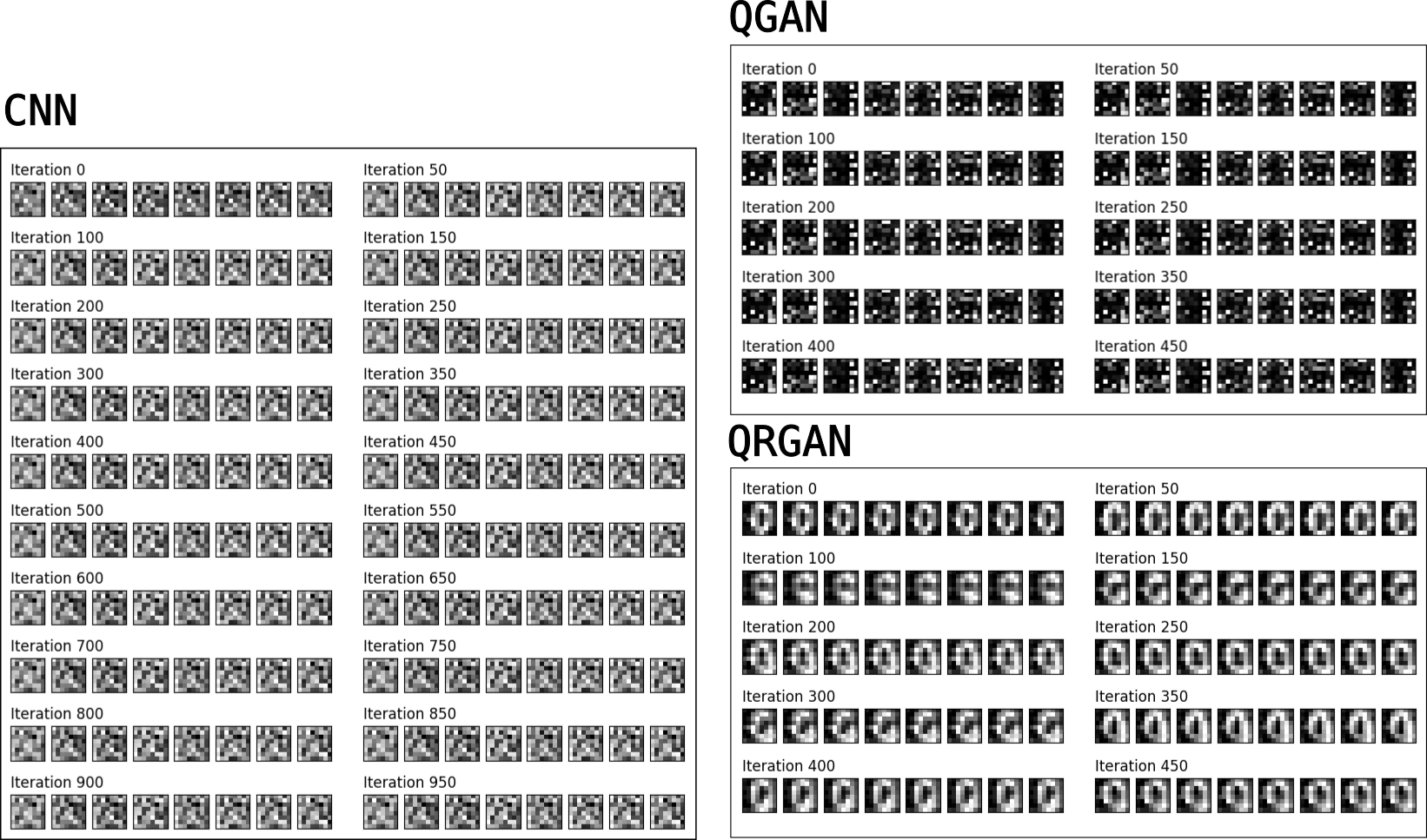}

\caption{
The generated picture of handwritten 0 by QRGAN, QGAN, and CNN using the loss function of Least Squares GAN as a loss function for the number of iterations.
The picture by QRGAN is generated by fixed noise and input.
}\label{ gen 0 l }

\end{figure}

On the other hand, we show the average of the loss function for 10 attempts of the discriminator and generator of QRGAN, QGAN and CNN for the number of epochs using the loss function of Least Squares GAN \cite{2016arXiv161104076M} in Fig . \ref{ comp 0 l }, and generated picture of 0 in Fig . \ref{ gen 0 l }, respectively.
The loss function of the generator on QGAN  declined to 0 faster than that of QRGAN  on average, the same as the case using cross-entropy as loss function, as shown in Fig.\ref{ comp 0 l }.
The loss function of the discriminator on QGAN depresses faster than that of QRGAN on average, too.
The values of the loss function converged at most in 100 iterations; hence, the generated pictures are closer to the input data than in the previous case.
However, generated pictures deviated from the shape of the aimed pictures according to Fig.\ref{ err }.
It is supposed to be because the QRGAN uses genuine data to generate fake data, and the similarity of the fake data to the genuine is maximized before the discriminator grows enough.

In addition, CNN could not improve the quality of the fake.

\begin{figure}

\includegraphics[scale=0.3]{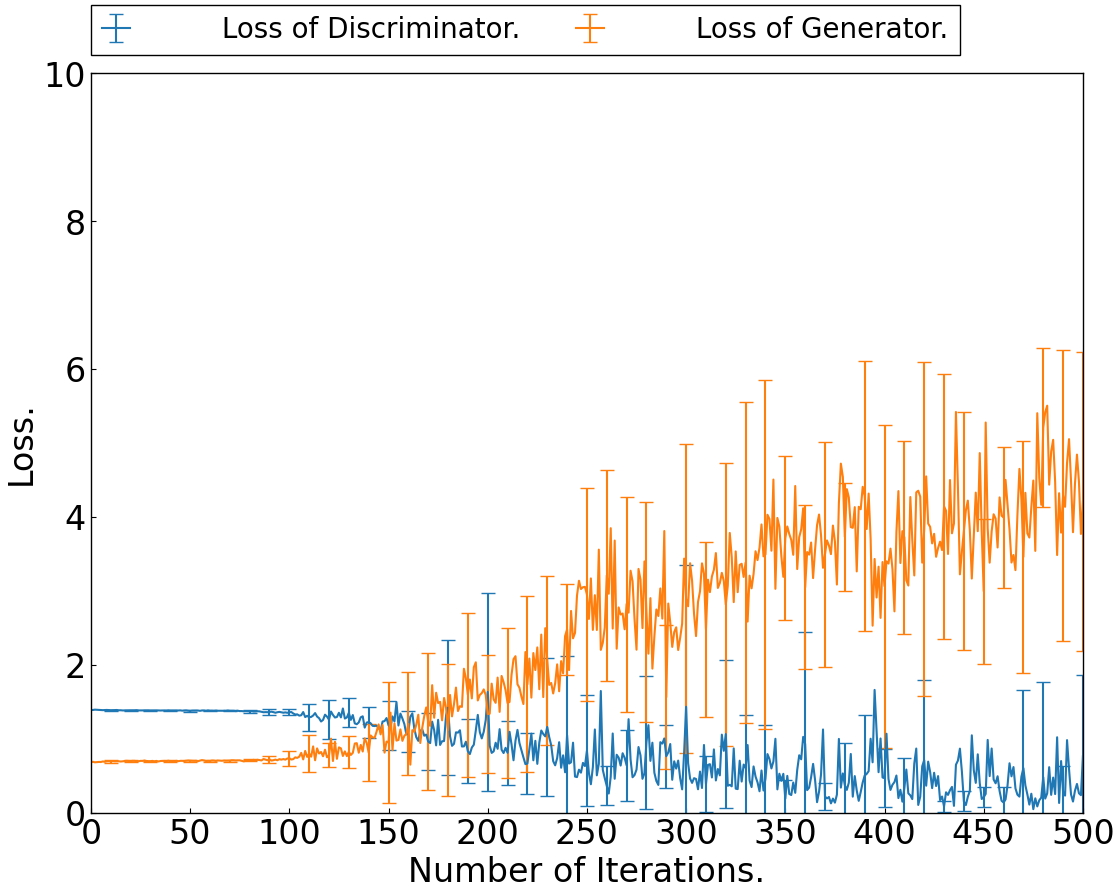}

\caption{ The number of iterations v.s. the average of the loss function of handwritten 0 to 9 of the discriminator and generator of QRGAN using cross-entropy as a loss function.   } \label{ comp s n }

\end{figure}

\begin{figure}

\includegraphics[scale=0.3]{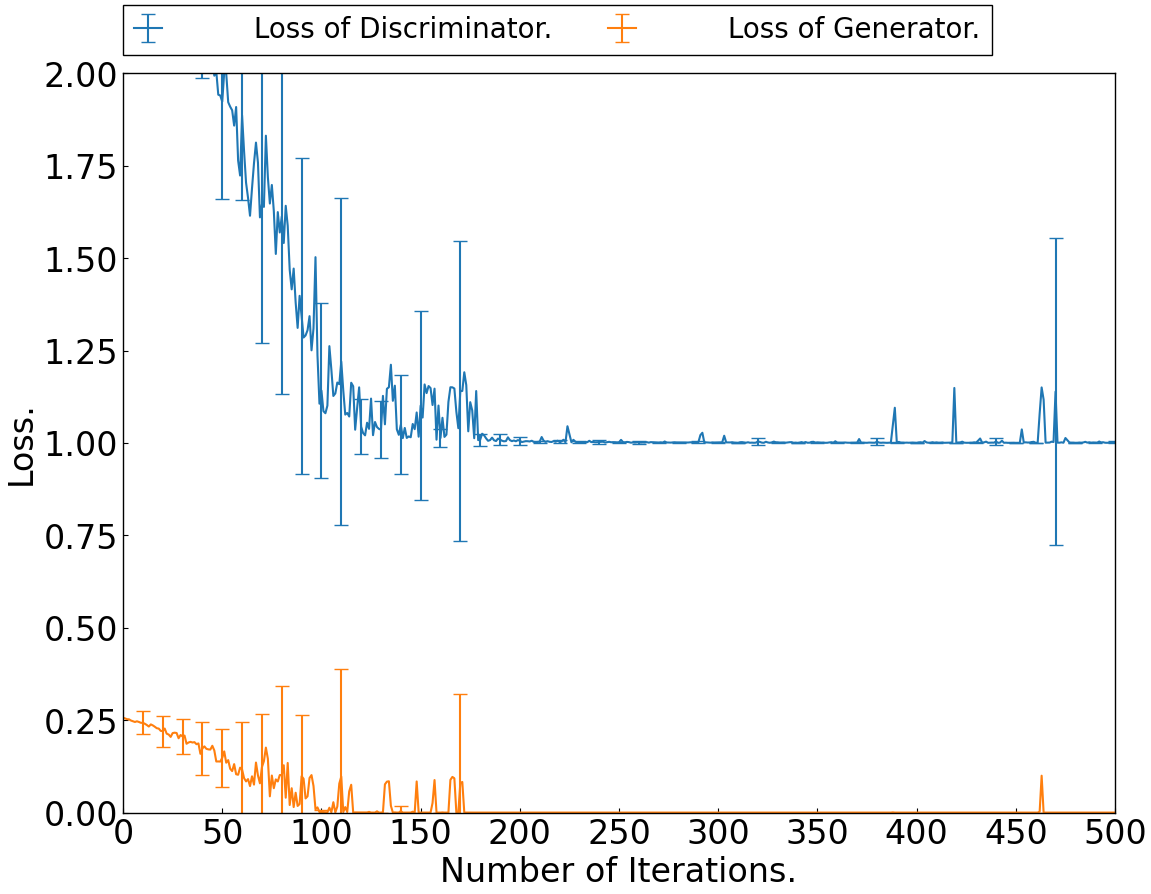}

\caption{  The number of iterations v.s. the average of the loss function of handwritten 0 to 9 of the discriminator and generator of QRGAN using the loss function of Least Squares GAN as a loss function.   }        \label{ comp n l }

\end{figure}

Secondly, we compared the result of QRGAN using the loss function of cross-entropy and that of Least Squares on handwritten single-digit numbers from 0 to 9.
The number of learned pictures is between 375 and 389.
We show the average of the loss function for processing each integer from 0 to 9 of the discriminator and generator of QRGAN for the number of epochs using cross-entropy in Fig . \ref{ comp s n }, and those using the loss function of Least Squares GAN in Fig . \ref{ comp n l }, respectively.
The loss function of the generator using cross-entropy rises as large as that of only 0, and the loss function of the discriminator depresses normally.
The loss function of the generator using the loss function of Least Squares GAN depresses more unstably than that of only 0, and the loss function of the discriminator also depresses unstably.

\begin{figure}

\includegraphics[scale=0.3]{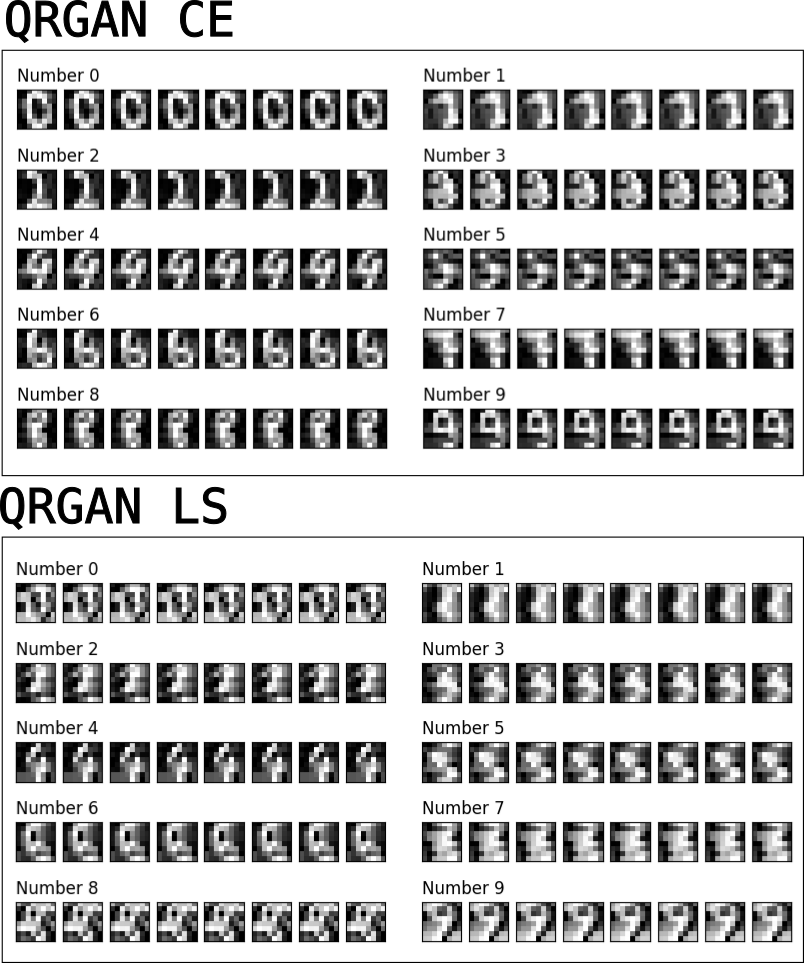}

\caption{ The generated picture of 0 to 9 by QRGAN using the loss function of Least Squares GAN and cross-entropy as a loss function for the 500th iterations.  The picture by QRGAN is generated by fixed noise and input. }   \label{ gena }

\end{figure}

On the other hand, we show the generated pictures of handwritten integers from 0 to 9 at the 500th iteration for processing each integer from 0 to 9 of QRGAN for the number of epochs using cross-entropy, and those using the loss function of Least Squares GAN in Fig.\ref{ gena }.

Generated 2, 3, 5, 7 using the loss function of Least Squares GAN is melted and hard to read.
It may be due to too fast growth of the discriminator, the same as the case of 0.
Generated numbers using cross-entropy as a loss function have a shape close to the input pictures, except for 3.
QRGAN may exhibit high accuracy for generating fakes for some kinds of pictures and low accuracy for other kinds of pictures.

\begin{figure} 
 
\includegraphics[scale=0.3]{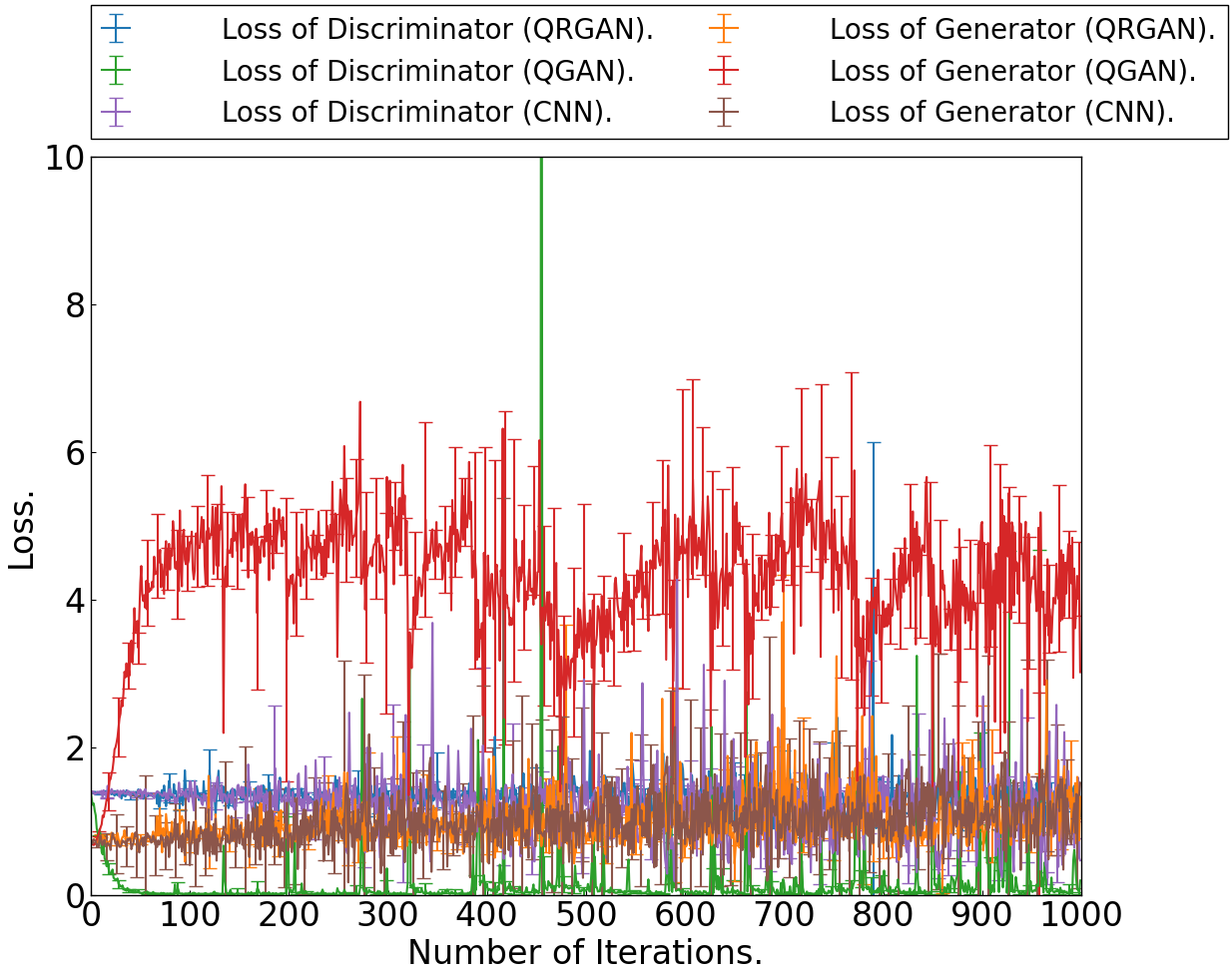} 

\caption{
The number of iterations v.s. the average of the loss function for 5 attempts of the discriminator and generator of QRGAN, QGAN, and CNN using cross-entropy as a loss function.
} \label{ pic s }  

\end{figure}

\begin{figure}

\includegraphics[scale=0.3]{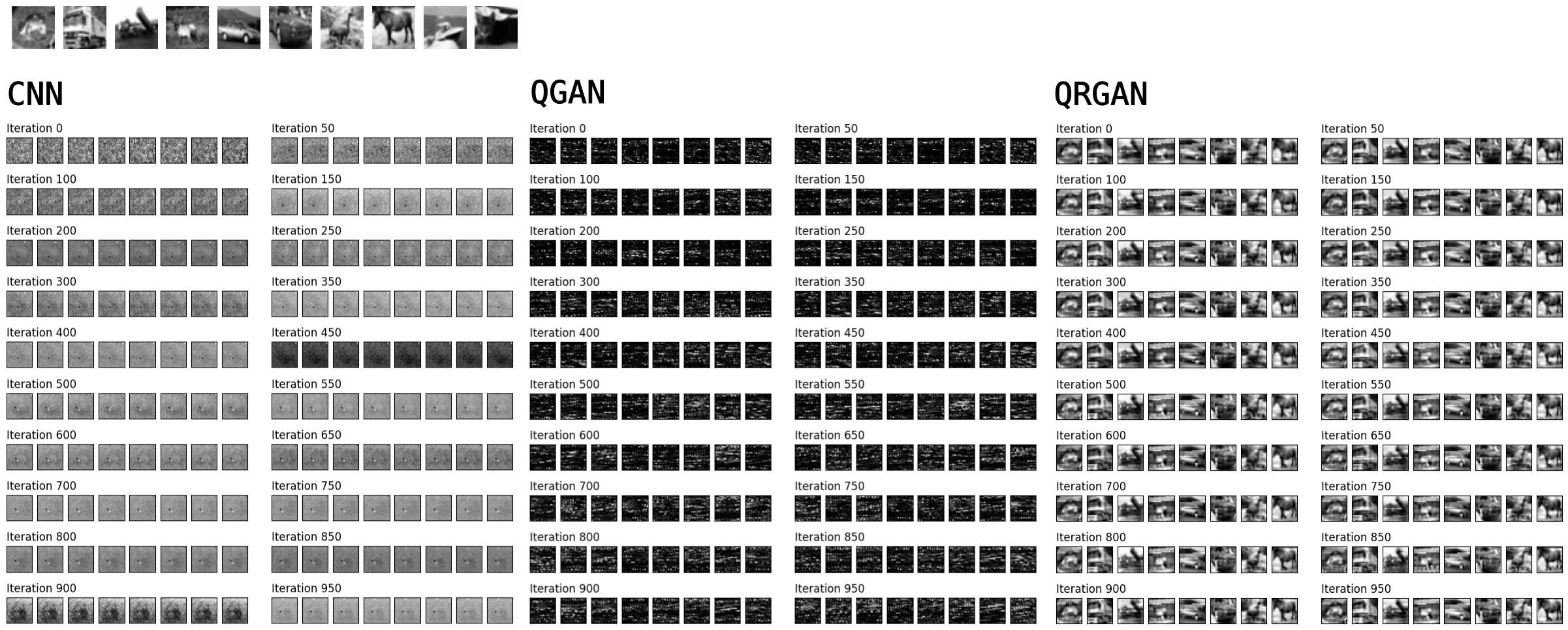} 

\caption{ 
The generated picture of the first 8 data by QRGAN, QGAN, and CNN using the cross-entropy as a loss function. 
The pictures in the upper right are the original pictures from 0 to 9 on the index.  }\label{ gen a }
  
\end{figure}

\begin{figure}

\includegraphics[scale=0.3]{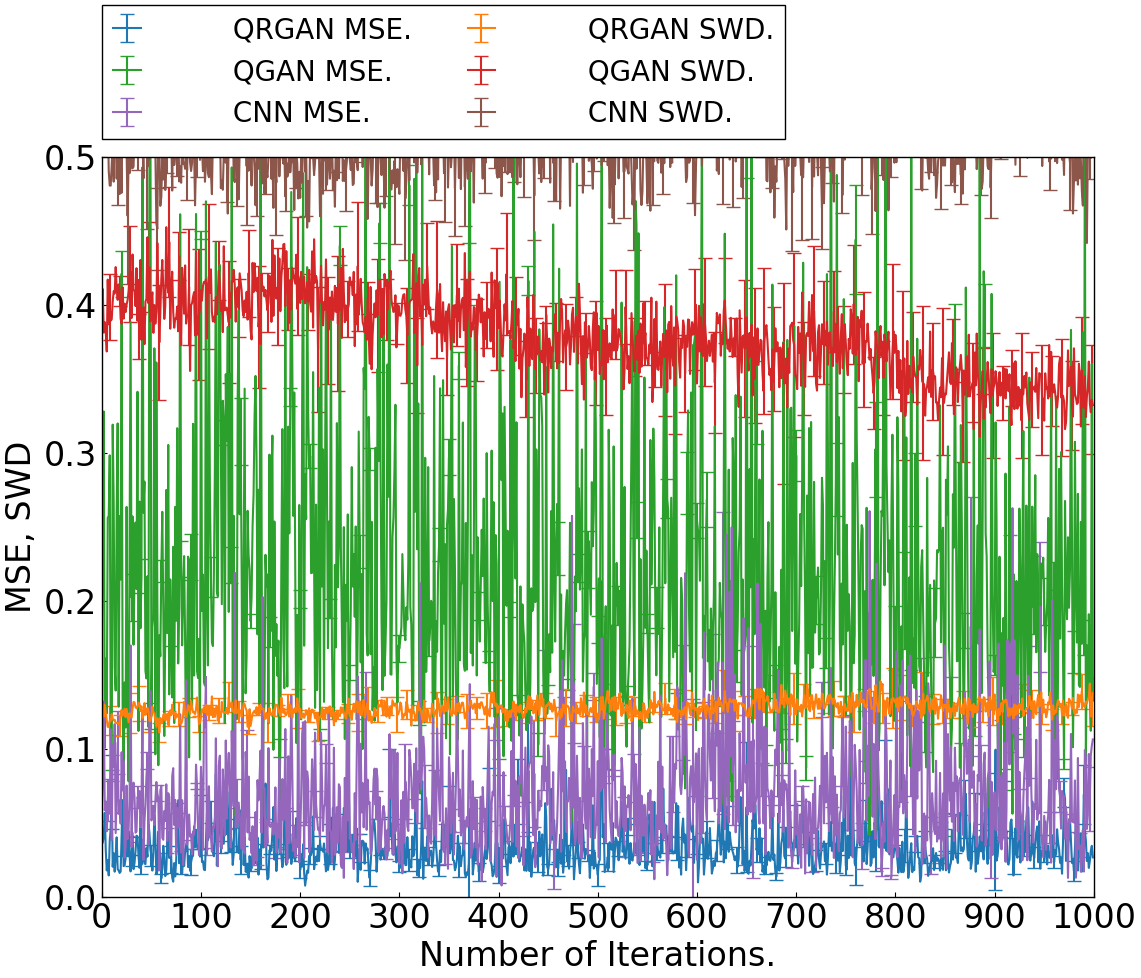}

\caption{  The number of iterations v.s. the average of the MSE and SWD for 5 attempts of QRGAN, QGAN, and CNN using cross-entropy as a loss function. } \label{ w d }
  
\end{figure}

Thirdly, we compared the result of generating a 16 $ \times $ 16 size picture on the CIFAR10 dataset.
We used only 1000 data from index 0. 
Both pictures are compressed from 32 $ \times $ 32 size original pictures into 22 $ \times $ 22 size and converted to gray scale.    
The number of qubits on QGAN is 8, and the size of the network on CNN is 8 times that of training on 0.     
We show the average of the loss function for 10 attempts of the discriminator and generator of QRGAN, QGAN, and CNN for the number of epochs using cross-entropy in Fig.\ref{ pic s }, and the generated picture of the first 8 pictures for fixed noise in Fig.\ref{ gen a }, respectively.          
 The average of QRGAN did not grow compared to QGAN and other cases of training, even though generated pictures have the smallest Sliced Wasserstein Distance (SWD) in all, also shown in Fig.\ref{ w d }.     
 We show the MSE and SWD of three methods in Fig.\ref{ w d }. 
SWDs are calculated by Python Optimal Transport \cite{JMLR:v22:20-451} on the first 8 pictures in the index of the original dataset.       
QRGAN has the smallest MSE and SWD of all ranges, and the SWD of QGAN declines gradually.
Besides, QRGAN has the smallest fluctuation in all methods because it uses genuine data to generate the fakes.
The accuracy of QRGAN declines little by iteration.
It is supposed to be because aimed data is learned only once for each, hence, repeated learning of aimed data may improve the accuracy.
Even though QGAN raises its accuracy by iteration, it will saturate because generating 8 aimed pictures is hard using only random noise.
The time for calculation of QRGAN is the largest on average, which is 8.6165 times larger than QGAN and 88.4877 times larger than CNN, respectively. 
If the network of CNN is larger, the time for calculation may be longer, and the accuracy may be far higher. 
Though the time for calculation on the simulation of QRGAN will never be smaller than that of CNN, the accuracy of QRGAN can not be surpassed by CNN. 
The time for the calculation of QRGAN can be shorter than that of QGAN if the number of shots can be smaller than that of QGAN keeping the accuracy.

Even generated pictures are further from genuine than those of single digits;  they have an error ratio below 0.1.
Rather, compression decreases the accuracy.

\section{Discussion}\label{5}

In this section, we discuss the reproduction property and the accuracy for the noisy input of QRGAN. 
 
\begin{figure}     
     
\includegraphics[scale=0.3]{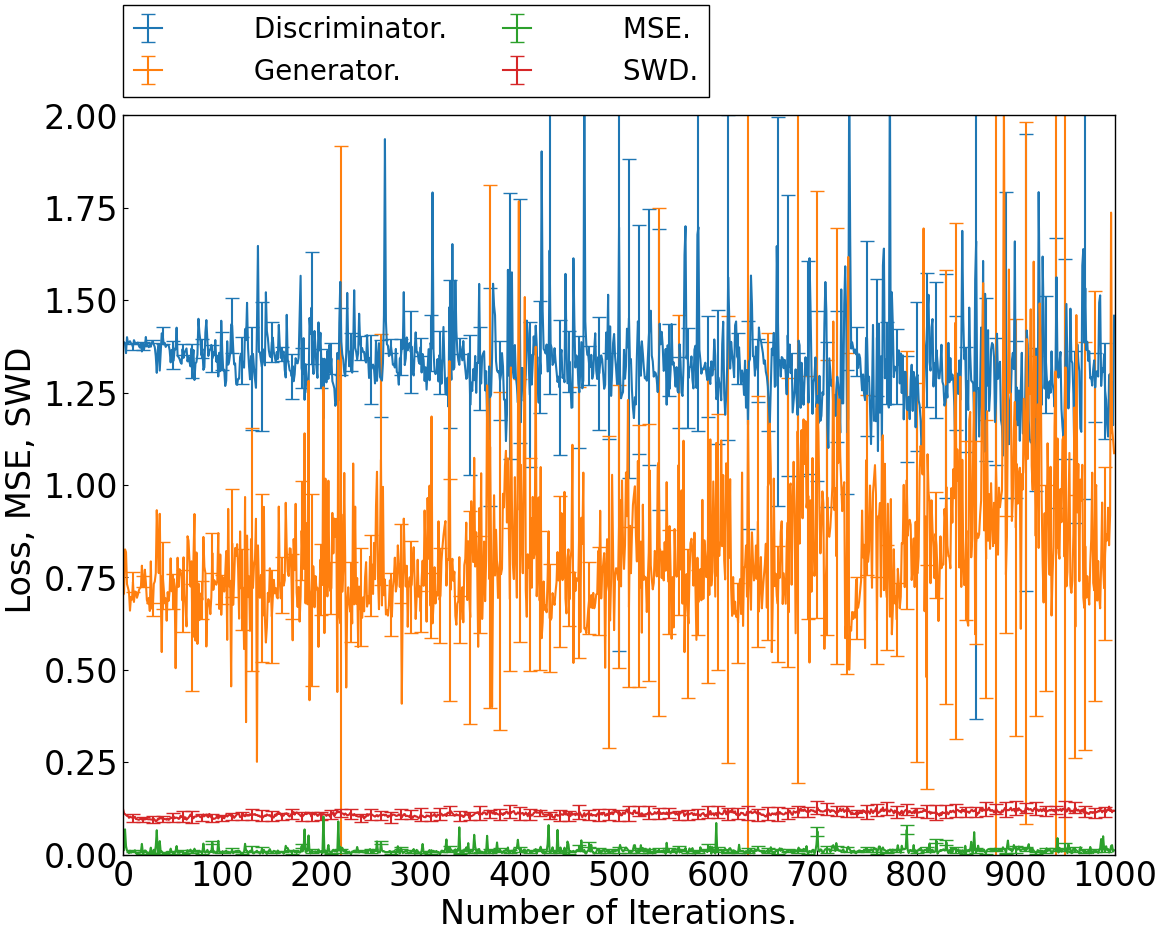}    
    
\caption{The number of iterations v.s. the average of the loss function, MSE, and SWD for 16 seeds of the discriminator and generator of QRGAN using cross-entropy as a loss function.} \label{ pic f s } 
     
\end{figure}

\begin{figure}     
 
\includegraphics[scale=0.3]{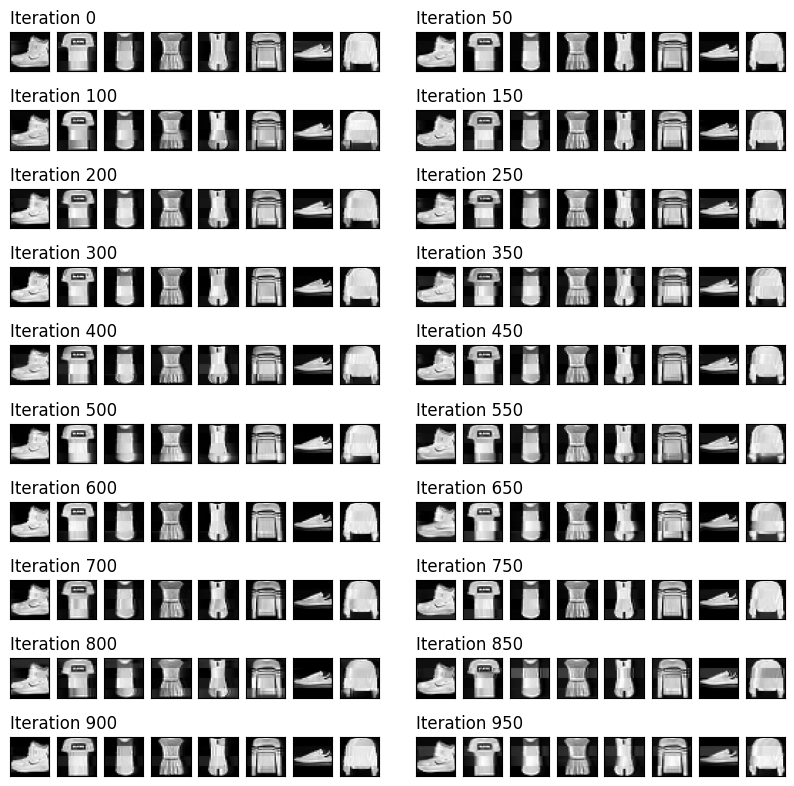}
 
\caption{The generated picture of the first 8 data for seed 17171 by QRGAN using the cross-entropy as a loss function.  }\label{ gen f a }
  
\end{figure}

First, we discuss the reproduction property of the output of QRGAN for data on Fashion-MNIST. 
We generated the first 8 data learning 1000 data using QRGAN , and calculated the loss functions of discriminator ,  the loss functions of generator , MSE and SWD for seed coefficient of random numbers $ 42 , 2 , 10 , 13 , 0 , 3407 , 7120 , 10000 , 11111 , 16384 , 17171 , 130000 , 14480 , 11668 , 500001 $ and $ 620000 $ and calculated t and p values by Bonferroni test.       
Data are all original size ( $ 28 \times 28 $ ).        
We show the average of the loss function, MSE, and SWD for 16 seeds of QRGAN for the number of epochs using cross-entropy in Fig.\ref{ pic f s }, and the generated picture of the first 8 pictures for fixed noise in Fig.\ref{ gen f a }, respectively.    
The generator learned and was enlarged by the progress of learning.    
The average and standard deviation of SWD are both small, the same as the result of CIFAR-10.     
According to the result of the Bonferroni test on SWD for all iterations, at least 9 seeds are clearly different from the given seed ( shown in Table . \ref { b t } ). 
The significance is $ 0.05 / 240 $, hence, the two seeds that have p value over 0 differ negligibly.       
Even though the SWD for different seeds has significant differences, the accuracy of QRGAN is the highest among QRGAN, QGAN, and CNN. 
       
\begin{table*}[h]  
   
 \caption{ The value of t and p values between two seeds on Fashion-MNIST of SWD of QRGAN using cross-entropy as a loss function.   }\label{ b t }      
   
\centering       
\resizebox{\textwidth}{!}{%
  
\begin{tabular}{c|c|c|c|c|c|c|c|c|c|c|c|c|c|c|c|c} \hline \hline                
 t value&42&2&10&13&0&3407&7120&10000&11111&16384&17171&130000&14480&11668& 500001&620000\\\hline  
42&0&-14.617&-15.658&0.7459&-5.0975&-4.2136&-11.7951&-3.1588&-6.061&-14.2671&-6.8997&-3.7044&-9.3736&-16.0357&-5.1444&-15.1776\\\hline    2&14.617&0&-2.5183&15.1065&9.2363&10.7525&2.6063&11.928&8.3838&0.5478&7.6038&9.8625&4.2003&-1.7685&9.3734&-0.19\\\hline     
10&15.658&2.5183&0&16.1046&10.8403&12.2101&4.8558&13.2559&10.0768&3.0388&9.3773&11.3839&6.2259&0.8625&10.9688&2.3935\\\hline 
13&-0.7459&-15.1065&-16.1046&0&-5.7367&-4.8954&-12.3249&-3.8664&-6.6879&-14.7659&-7.5142&-4.3338&-9.9173&-16.4972&-5.7902&-15.6643\\\hline   
0&5.0975&-9.2363&-10.8403&5.7367&0&1.0998&-6.5486&2.1765&-0.9038&-8.8147&-1.7045&1.086&-4.485&-10.7689&0.0221&-9.6402\\\hline  
3407&4.2136&-10.7525&-12.2101&4.8954&-1.0998&0&-7.929&1.1208&-2.0525&-10.3409&-2.8935&0.0917&-5.6911&-12.2951&-1.092&-11.2248\\\hline 
7120&11.7951&-2.6063&-4.8558&12.3249&6.5486&7.929&0&9.0636&5.6889&-2.0978&4.91&7.3034&1.7126&-4.2958&6.651&-2.8531\\\hline 
10000&3.1588&-11.928&-13.2559&3.8664&-2.1765&-1.1208&-9.0636&0&-3.1462&-11.5324&-3.9989&-0.9123&-6.7357&-13.4478&-2.1842&-12.4396\\\hline 
11111&6.061&-8.3838&-10.0768&6.6879&0.9038&2.0525&-5.6889&3.1462&0&-7.9485&-0.8032&1.9491&-3.6588&-9.9454&0.9377&-8.7709\\\hline 
16384&14.2671&-0.5478&-3.0388&14.7659&8.8147&10.3409&2.0978&11.5324&7.9485&0&7.1568&9.4638&3.7356&-2.3247&8.95&-0.7533\\\hline 
17171&6.8997&-7.6038&-9.3773&7.5142&1.7045&2.8935&-4.91&3.9989&0.8032&-7.1568&0&2.7109&-2.9163&-9.1887&1.7487&-7.9736\\\hline 
130000&3.7044&-9.8625&-11.3839&4.3338&-1.086&-0.0917&-7.3034&0.9123&-1.9491&-9.4638&-2.7109&0&-5.3102&-11.32&-1.0774&-10.2488\\\hline 
14480&9.3736&-4.2003&-6.2259&9.9173&4.485&5.6911&-1.7126&6.7357&3.6588&-3.7356&2.9163&5.3102&0&-5.7757&4.5554&-4.4636\\\hline 
11668&16.0357&1.7685&-0.8625&16.4972&10.7689&12.2951&4.2958&13.4478&9.9454&2.3247&9.1887&11.32&5.7757&0&10.9171&1.6214\\\hline 
500001&5.1444&-9.3734&-10.9688&5.7902&-0.0221&1.092&-6.651&2.1842&-0.9377&-8.95&-1.7487&1.0774&-4.5554&-10.9171&0&-9.7885\\\hline 
620000&15.1776&0.19&-2.3935&15.6643&9.6402&11.2248&2.8531&12.4396&8.7709&0.7533&7.9736&10.2488&4.4636&-1.6214&9.7885&0\\\hline \hline 
p value&42&2&10&13&0&3407&7120&10000&11111&16384&17171&130000&14480&11668&500001&620000\\\hline  
42&1&0&0&0.4558&0&0&0&0.0016&0&0&0&0.0002&0&0&0&0\\\hline   
2&0&1&0.0119&0&0&0&0.0092&0&0&0.5839&0&0&0&0.0771&0&0.8493\\\hline  
10&0&0.0119&1&0&0&0&0&0&0&0.0024&0&0&0&0.3885&0&0.0168\\\hline 
13&0.4558&0&0&1&0&0&0&0.0001&0&0&0&0&0&0&0&0\\\hline   
0&0&0&0&0&1&0.2715&0&0.0296&0.3662&0&0.0884&0.2776&0&0&0.9823&0\\\hline 
3407&0&0&0&0&0.2715&1&0&0.2625&0.0402&0&0.0039&0.9269&0&0&0.275&0\\\hline 
7120&0&0.0092&0&0&0&0&1&0&0&0.036&0&0&0.0869&0&0&0.0044\\\hline 
10000&0.0016&0&0&0.0001&0.0296&0.2625&0&1&0.0017&0&0.0001&0.3617&0&0&0.0291&0\\\hline 
11111&0&0&0&0&0.3662&0.0402&0&0.0017&1&0&0.4219&0.0514&0.0003&0&0.3485&0\\\hline 
16384&0&0.5839&0.0024&0&0&0&0.036&0&0&1&0&0&0.0002&0.0202&0&0.4513\\\hline 
17171&0&0&0&0&0.0884&0.0039&0&0.0001&0.4219&0&1&0.0068&0.0036&0&0.0805&0\\\hline 
130000&0.0002&0&0&0&0.2776&0.9269&0&0.3617&0.0514&0&0.0068&1&0&0&0.2814&0\\\hline  
14480&0&0&0&0&0&0&0.0869&0&0.0003&0.0002&0.0036&0&1&0&0&0\\\hline 
11668&0&0.0771&0.3885&0&0&0&0&0&0&0.0202&0&0&0&1&0&0.1051\\\hline 
500001&0&0&0&0&0.9823&0.275&0&0.0291&0.3485&0&0.0805&0.2814&0&0&1&0\\\hline 
620000&0&0.8493&0.0168&0&0&0&0.0044&0&0&0.4513&0&0&0&0.1051&0&1\\\hline  
\end{tabular}                   
}%
\end{table*}

\begin{figure}
  
\includegraphics[scale=0.3]{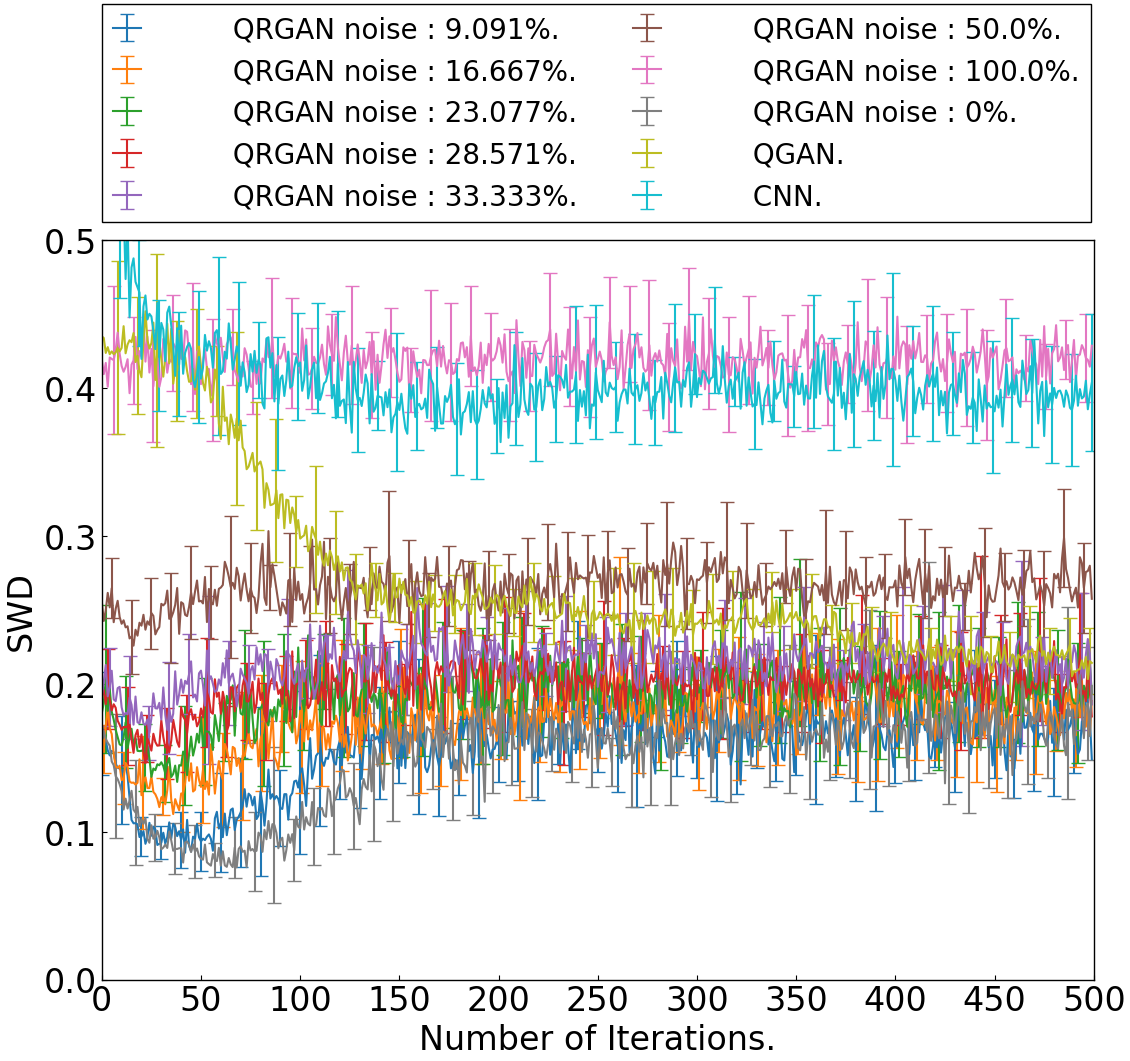}  
\caption{ 
The number of iterations v.s. the SWD for 10 attempts of the discriminator and generator of QRGAN for various ratios of intensity of noise on input, QGAN, and CNN using cross-entropy as a loss function. 
Error bars indicate the standard deviation from the average, and each bar is sampled every 10 points, moving 1 point from the previous data.
} \label{ d s }
  
\end{figure}

\begin{figure} 

\includegraphics[scale=0.3]{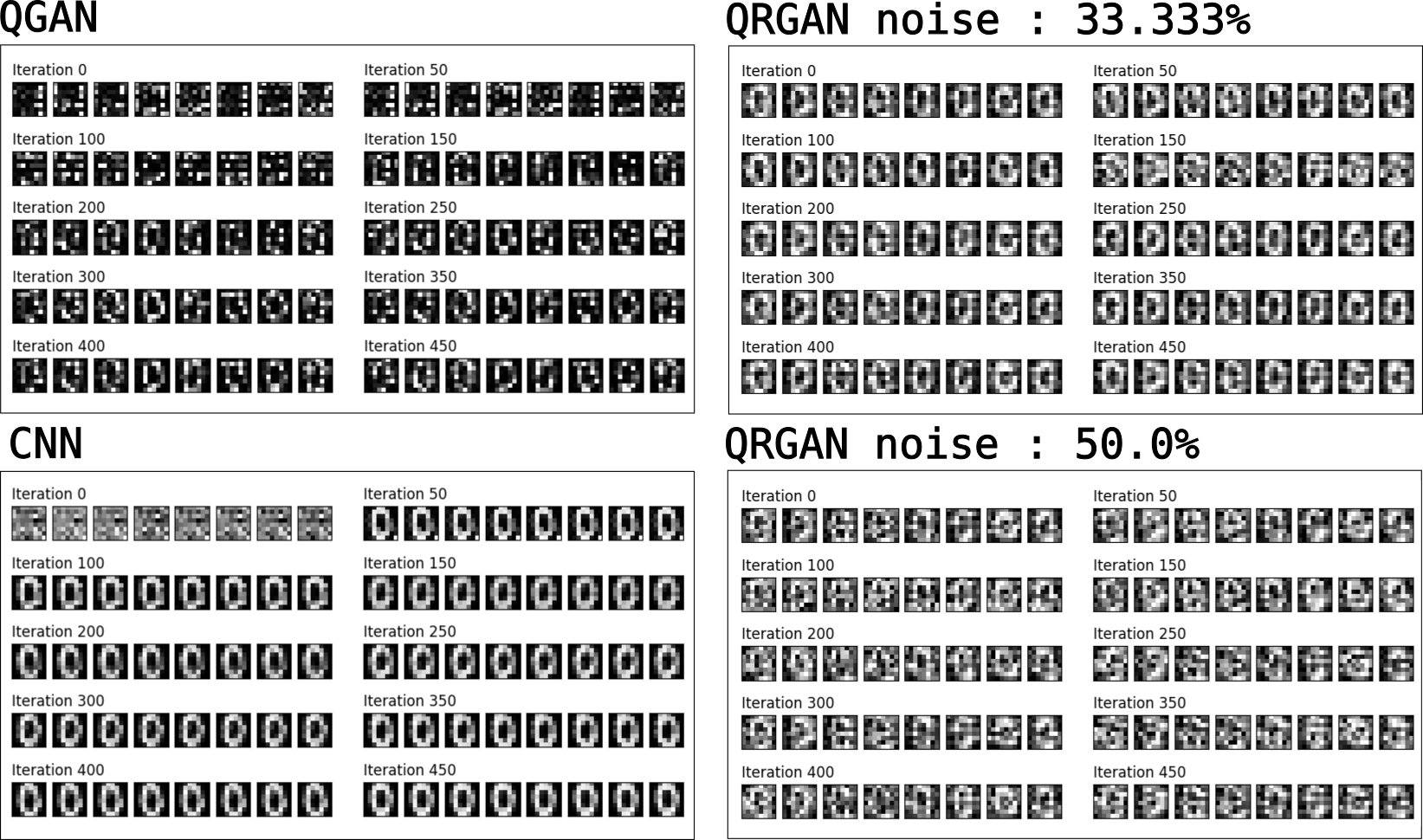} 

\caption{
The generated picture of the chosen 8 data by QRGAN, QGAN, and CNN using the cross-entropy as a loss function.
}\label{ d s p }

\end{figure}

All data of QRGAN uses genuine data for input.   
This fact does not matter for improving the accuracy of QRC. 
Rather, it is the advantage that input data can guide the generated fakes. 
We are interested in the ability to guide the generated data; therefore, we surveyed it.   
We show the SWD on randomly picked 8 data in 375 data of 0 in Optdigits in advance for QRGAN whose input data are mixed with various intensities of noise, QGAN, and CNN  in Fig . \ref{ d s }.  
QRGAN with noisy input exhibited the bottom for the number of iterations, except for pure noise input.
In case the intensity ratio of noise is below 1 / 3, QRGAN is more accurate than QGAN and CNN  for the range less than or equal to 500 iterations. 
Besides, generated pictures reduce noise as shown in Fig.\ref{ d s p }. 
QRGAN can be used for noise reduction of data by repeating the replacement of input data to generated fake data.
The SWD of the method can be improved by this procedure, even if it is worse than those of other methods, just like Feedback-driven QRC \cite{PRXQuantum.5.040325}.
 
The accuracy of QRGAN is higher than that of other QRC \cite{Nerenberg:25} and QGAN \cite{2021PhRvP..16b4051H,math12233852} methods for the picture of 0 and the gray-scale pictures.   
However, QRGAN can not treat the large amount of data like QINR-based QGAN \cite{ma_quantum_2025} and Photon-QuaRC \cite{Nerenberg:25}; hence, the method to train with a larger amount of data is the next problem.
Besides, the result is the case that we used the state vectors for readout for QRGAN and QGAN, not the statistics of the shots.  
QGAN is weak for noise due to the required $ 6 N_q $ control gates compared to QRGAN, because QRGAN requires $ ( N_q - 1 ) N_q $ control gates, the number of qubits is not necessary for large-scale calculation, and QRGAN uses mixed states. 
The number of terms of QRGAN can be saved by the number $ ( N_q - 1 ) N_q $.   
QRGAN can be more accurate than QGAN on real devices, even though the time for calculation is nearly ten times larger than QGAN.                    
The method to train by a small number of shots as accurately as the result using state vectors is also the next problem.

\section{Concluding remarks}\label{7}    
         
In this paper, we confirmed that QRGAN is more accurate than QGAN, CNN, and QRC for single handwritten digits and small pictures.
Quantum Reservoir Computers can be the generator of GAN, and they require no additional qubits for extending the data length; hence, they will be helpful in large-scale data processing by quantum computers accurately. 
However, training methods for larger amounts of data accurately and fast remained as the following problem.
Besides, our simulations use genuine data for QRC; hence, but for them, the accuracy declines. 
Establishing a method to calculate accurately using noisy input is a problem, too.
Data from this paper is all the result of simulation; hence, simulation on real devices is also a future plan.

\bibliography{mainwakaura2}


\end{document}